\documentclass[pre,twocolumn,notitlepage,longbibliography,amsmath,amssymb,floats,superscriptaddress,nofootinbib,10pt]{revtex4-2}
\usepackage{graphicx,color}
\usepackage{amsmath}
\usepackage{amsfonts, amsbsy}
\usepackage{amssymb}
\usepackage{eqnarray}
\usepackage{bm}
\usepackage{blkarray}
\usepackage{mathtools}
\usepackage{psfrag}
\usepackage{caption}
\usepackage{subcaption}
\usepackage{multirow}
\usepackage{textcomp}
\usepackage{units}
\usepackage{lipsum}
\usepackage[title]{appendix}
\usepackage{soul}
\usepackage{titlesec}
\usepackage{times}
\usepackage{enumitem}
\usepackage{soul}
\usepackage[sort&compress]{natbib}
\usepackage[breaklinks,colorlinks = true,linkcolor = blue,urlcolor  = blue,citecolor = blue,anchorcolor = blue]{hyperref}
\usepackage{float}

\begin{document}
\title{Generalizing odd elasticity theory to odd thermoelasticity for planar materials}
\author{Martin Ostoja-Starzewski}
\email{martinos@illinois.edu}
\affiliation{Department of Mechanical Science \& Engineering, University of Illinois at Urbana-Champaign Urbana, IL 61801, USA}
\affiliation{Beckman Institute, University of Illinois at Urbana-Champaign Urbana, IL 61801, USA}
\author{Piotr Sur\'{o}wka}
\email{piotr.surowka@pwr.edu.pl}
\affiliation{Institute of Theoretical Physics, Wroc\l{}aw University of Science and Technology, 50-370, Wroc\l{}aw, Poland}
\date{\today}

\begin{abstract}
We generalize the odd elasticity of planar materials to thermoelasticity,
admitting spatially inhomogeneous properties. First, we show that for active systems breaking Onsager relations thermal evolution is given by an odd generalization of the Maxwell-Cattaneo relation. Next three different heat
conduction models of odd solids are considered leading, respectively, to a classical
coupled thermoelasticity with Fourier law, thermoelasticity with
relaxation times of the Maxwell-Cattaneo type, and thermoelasticity with two
relaxation times. Governing equations are established in terms of either
displacement-temperature pair, stress-heat flux pair, or stress-temperature
pair. Next, we establish a form of the stiffness tensor, ensuring its
inversion to a compatibility tensor, and write equations of elasticity in
the presence of eigenstrains, such as thermal strains, where we find that
the stress field remains unchanged for a specific additive change of the
compliance tensor field. This so-called stress invariance gives an
equivalence class of a wide range of odd materials with different values of
material properties. Effectively, within each class, the elastic compliances
may be modified by a field linear in the plane without affecting the stress
field. Finally, we study hydrodynamic modes in an odd thermoelastic solid with Fourier heat conduction and argue that contrary to even elastic solids, the temperature can affect both dilatational and shear waves. We present odd corrections to sound attenuation and diffusion coefficients.
 \end{abstract}

\maketitle

\section{Introduction} Thermoelasticity, the interdisciplinary study that converges the principles of thermodynamics with those of elasticity, is a cornerstone in the realm of continuum mechanics. As a field of inquiry, it bridges the gap between thermal and mechanical behavior in materials. Understanding thermoelastic phenomena is crucial for a host of applications ranging from industrial processes, aerospace engineering, and material science to cutting-edge research in nanotechnology and biological systems.

As the demands for advanced materials with specific properties grow, be it in extreme temperatures, pressures, or other challenging conditions, the predictive power of thermoelastic models becomes increasingly significant. Traditional elasticity theories may fall short when subjected to varying temperature fields, often resulting in inaccurate predictions and undesirable outcomes in applications such as turbine blade design, thermal barrier coatings, or nuclear reactor construction. Herein lies the indubitable importance of thermoelasticity—it provides a more comprehensive model, capturing the intricate interplay between thermal and elastic effects, to achieve better accuracy and reliability in predictive analysis and design.

Moreover, the theory of thermoelasticity finds applications in the real-time analysis of stress and strain in nanostructures, high-speed machinery, and other systems where both mechanical and thermal effects cannot be ignored. With the advent of high computational capabilities, solving complex thermoelastic problems is becoming increasingly feasible, thereby opening new avenues for innovation and application.

Traditionally, thermoelasticity has been employed to understand how conventional materials respond to thermal and mechanical stimuli. In active matter, this traditional framework needs to be extended to account for the internal energy sources that drive system behavior. Unlike passive systems, active matter is characterized by a constant input of energy at the microscopic level, leading to macroscopic patterns of motion and deformation. This inherently makes the thermoelastic description of active matter far more complex but also far more intriguing.

The recent few years have witnessed the
development of odd elasticity, a theoretical framework for elastic materials
that do not store energy in the same way as hyperelastic materials do 
\cite{Scheibner-2020,banerjee_active_2021,lier_passive_2022,Surowka-2022} (for a review see \cite{Fruchart}). Various physical systems display such responses, typically due to the
breakdown of Maxwell-Betti reciprocity. At this point, new challenges arise.
First, to extend the framework of odd elasticity to non-isothermal and/or
non-adiabatic behaviors. This is considered here with three thermoelastic
models: classical coupled thermoelasticity with Fourier law,
thermoelasticity with relaxation times of Maxwell-Cattaneo type, and
thermoelasticity with two relaxation times. While the latter two models are
fundamentally different from one another, they can be simplified to the classical coupled one when the relaxation times are set to zero. The goal is to write down the governing equations
admitting, in general, spatially inhomogeneous properties. The next question
concerns the possibility of invariance of the stress field in an odd
thermoelastic two-dimensional (planar) material. This problem of a so-called
''Cherkaev-Lurie-Milton (CLM) shift in a compliance tensor field" \cite{CLM} is tackled in a more
general setting of eigenstrains. We determine an equivalence class of a wide range of odd materials in which the elastic compliances may be modified without affecting
the stress field.

\section{Irreversible thermodynamics} We construct the first odd extension of thermoelasticity by using the language of extended irreversible thermodynamics. In active systems the concept of effective temperature is introduced as an analogous quantity to the conventional temperature in equilibrium systems \cite{loi_effective_2008}. It serves as a tool to describe the energy distribution or the degree of agitation of the particles in the system, despite the lack of true thermal equilibrium. However, it is important to note that the effective temperature is not a true thermodynamic temperature. It can vary significantly from the actual thermodynamic temperature of the environment and can even depend on the specific properties of the particles or the type of measurement being made. Moreover, the concept of temperature out-of-equilibrium is not unique. Therefore, whenever we talk about an effective temperature of an active solid, we assume a measurement scheme has been prefixed to measure the effective temperature of the active solid and the temperature field is given by the thermometer corresponding to the scheme.

In the initial development of thermoelasticity for active solids, activity is incorporated by violating Onsager's reciprocity principle. We start by writing the differential for the entropy. We assume that it is a function of the energy, symmetrized strain, and heat current \cite{jou_extended_2001}
\begin{equation}
s=s(\epsilon,\varepsilon_{ij},q_i),
\end{equation} 
where $\varepsilon
_{kl}=u_{(k,l)}$. This is analogous to viscoelasticity, where the stress or momentum current also contributes to entropy. Taking the divergence one gets
\begin{equation}
ds=\frac{d \epsilon}{T}+\frac{\partial s}{\partial \varepsilon_{ij}}d \varepsilon_{ij}+\frac{\partial s}{\partial q_k} dq_k.
\end{equation} 
We impose the second law of thermodynamics
\begin{equation}
\Delta _s= \dot s+\nabla _i J^s_i\geq 0,
\end{equation} 
where $J^s$ is the entropy current, and supplement the system with conservation laws
\begin{align}
      &  \rho \ddot{u}_{i}   + \partial_{j} t_{ i j  }     =   0,
 \label{momentumcons}      
 \\ 
 &      \dot{\epsilon} +  \partial_j  q_{j}  + \dot \varepsilon_{ij} t_{ i j  }     =   0,   \label{energycon} \\
 &  \dot{ \rho}      =  0 \label{masscon}.
\end{align}
Here $\rho$ is the mass density, which we set to unity. Using the conservation laws, after some algebra, we find
\begin{equation}
\partial_i (J^s_i-q^i/T )+q_i \nabla_i (1/T)+ \left( \frac{\partial s}{\partial \varepsilon_{ij}}+t_{ij} \right) \dot \varepsilon_{ij}+\lambda_i \dot q _i\geq 0,
\end{equation} 
where we have defined $\lambda_i\equiv\partial s/\partial q_i$. The positivity of the entropy generation leads to $J^s_i=q_i/T$
, $ \partial s/\partial u_{ij}=-t_{ij}$.
We are left with
\begin{equation}
\Delta _s=q_i \nabla_i (1/T)+ \lambda_i \dot q _i\geq 0.
\end{equation} 
We now assume, in the linear regime, that
\begin{equation}
 \lambda_i =-\alpha_{ij} q_j ,
\end{equation} 
for some phenomenological tensor $\alpha_{ij}$. When $\alpha_{ij}$ is positive the system is bounded from below with a well-defined equilibrium state. Although we have active matter in mind we still assume that our system is bounded from below. Similarly, we impose
\begin{equation}
\nabla_i (1/T)+ \alpha_{ij}  \dot q_{j} = \gamma_{ij} q_j,
\end{equation} 
from which relation we obtain
\begin{equation}
\tau _{ij} \dot q_j=-k _{ij} \nabla_j T- q_i,
\end{equation} 
where $\tau_{ij}=\alpha_{ik}\gamma_ {kj}^{-1}$, $k_{ij}=\gamma _{ij}^{-1}/T^2$. In passive systems, the above equation follows from the positivity of the entropy production. Note that in the universe of active matter, this relation is not the most general. However, it is not our goal to introduce as many mechanisms for activity as possible, but rather have a minimal set-up that leads to non-trivial physical phenomena. In this case, we assume that activity is introduced by breaking Onsager relations. For even passive materials ($\alpha_{ij}=\alpha \delta_{ij}$, $\gamma_{ij}=\gamma_{ji}=\gamma \delta_{ij}$) this reduces to the isotropic, even Maxwell-Cattaneo relation
\begin{equation}
\tau  \dot q_i=-\lambda \nabla_i T- q_i,
\end{equation} 
where $\tau =\alpha \gamma$ and $\lambda = \gamma/T^2$. However, for odd active thermoelastic materials, we can have a generalized relation, with odd components $\tau_{\text{odd}}$ and  $k_{\text{odd}}$. $k_{\text{odd}}$ is an active component of heat conductivity in odd materials. $\tau_{\text{odd}}$ is a relaxation time of the odd heat flux. Its value is independent of the even component. Odd and even heat propagation have a different physical origin, therefore they do not need to relax in the same way. This is analogous to odd relaxations in odd viscoelasticity \cite{banerjee_active_2021}. The necessity to modify the Maxwell-Cattaneo law for odd active materials that do not obey Onsager symmetry is one of the central results of this paper. After presenting this basic model that highlights odd effects in thermoelasticity of active solids, we will delve into more comprehensive models. These models will not only address the violation of Onsager relations but also incorporate odd elasticity, as well as body forces and heat rate.

\section{Thermoelasticity with parabolic or hyperbolic heat conduction} 

The Fourier model of heat conduction works very well when the signal propagation speed does not need to be accounted for; this is the basis of classical coupled (C-C) thermoelasticity. At sufficiently high modulation frequencies, the time delay between the temperature gradient and heat flux needs to be considered, and this results in a Maxwell-Cattaneo model with one relaxation time, which is the basis of the Lord-Shulman (L-S) thermoelasticity, e.g. \cite{salazar_energy_2006,caceres_finite-velocity_2020}. By contrast, the Green-Lindsay (G-L) thermoelasticity retains the Fourier law but accounts for two time delays in coupled field phenomena: one between the stress and stress-temperature tensor, and another between the entropy and the specific heat. Thus, with two relaxation times G-L offers a theory which is intermediate between C-C and L-S theories.

In the subsequent analysis, we focus on the thermal properties of odd elastic solids. This activity stems from breaking Maxwell-Betti reciprocity relations. Incorporation of thermal strains and stresses in elasticity leads to three basic models of thermoelasticity that generalize the above classic models to the realm of odd solids:

(i) Heat conduction based on the Fourier law 
\begin{equation} \label{Fourier law}
q_{i}=-k_{ij}T,_{j}, 
\end{equation}
where $k_{ij}$\ is the thermal conductivity tensor (such that $k_{ij}T,_{i}T,_{j}>0$). Depending on the particular constitution of the odd
material, the Onsager symmetry of $k_{ij}=k_{ji}$\ may be broken.

(ii) Heat conduction involving one relaxation time $\tau $ of the
Maxwell-Cattaneo law 
\begin{equation}
q_{i}+\tau\dot{q}_{i}=-k_{ij}T,_{j},  \label{M-C law}
\end{equation}
where, just as in (12), the overdot stands for the material time derivative \cite{Christov-Jordan}. As pointed out in that reference, this model is
useful in various materials, typically polymeric, including the living \cite{JRSI-2023} or dead \cite{JRSI-2019} soft bio-tissues where $\tau $ may be on the order of one or several seconds. The resulting wave-type heat propagation is referred to as the 
\textit{second sound} as opposed to the usual elastic waves (\textit{first sound}). Also in this model, the Onsager symmetry may be broken. Hereinafter, we switch to the standard tensor notation for a gradient: a comma followed by the index of coordinate.

(iii) Heat conduction involving two relaxation times $\tau _{0}$\ and $\tau
_{1}$ \cite{IO-S}. This model provides an alternative formulation of
wave-type heat conduction. Again one can further consider generalizations of this model with broken Onsager symmetry.

Since elastodynamics itself is a hyperbolic theory, the first of these
models lead to a coupled hyperbolic-parabolic thermoelastic system, while
(ii) and (iii) are purely hyperbolic, albeit each different in character.

In cases (i) and (ii), the thermoelastic constitutive law reads $%
\varepsilon _{ij}=S_{ijkl}\sigma _{kl}+A_{ij}\Delta T$, where $\varepsilon
_{kl}=u_{(k},_{l)}$\ is the strain with $\left( k,l\right) $\ denoting
symmetrization on indices $i$ and $j$, $u_{i}$\ is the displacement; $%
S_{ijkl}$ is the compliance tensor (to be discussed below), and $\sigma
_{kl} $\ is the Cauchy stress tensor. Also, $A_{ij}$\ ($=-S_{ijkl}M_{kl}$)\
is the thermal expansion tensor and $\Delta T$\ is the\ temperature change
from $T_{0}$. The thermoelasticity field equations can be compactly grasped
in one system of two coupled equations for the displacement-temperature pair:%
\begin{equation}
\begin{array}{c}
(C_{ijkl}u_{k},_{l}),_{j}+\left( M_{ij}T\right) ,_{j}+b_{i}=\rho \ddot{u}_{i}
\\ 
\left( k_{ij}T,_{j}\right) ,_{i}-c_{e}\left( \dot{T}+\tau \ddot{T}\right)
+T_{0}\left[ M_{ij}(\dot{u}_{i}+\tau \ddot{u}_{i})\right] ,_{j}=-r-\tau \dot{%
r},%
\end{array}
\label{u-T pair}
\end{equation}%
Here $C_{ijkl}=S_{ijkl}^{-1}$\ is the elasticity (or stiffness) tensor which
has minor symmetries ($C_{ijkl}=C_{jikl}=C_{ijlk}$) but not the major one; $%
M_{ij}$\ is the stress-temperature tensor without symmetry ($M_{ij}\neq
M_{ji}$); $T$ is the absolute temperature, $T_{0}$ is the effective reference
temperature, $b_{i}$\ is the body force per unit volume, $\rho $\ is the
mass density, $r$ is the heat produced per unit time and unit mass, and $%
c_{e}>0$ is the specific heat at zero strain. When $\tau >0$, we have the
so-called Lord-Shulman (L-S) thermoelasticity \cite{Lord-Shulman} and, when $%
\tau =0$, the classical coupled (C-C) thermoelasticity is recovered.

On the other hand, case (iii) is the basis of the Green-Lindsay (G-L)
thermoelasticity \cite{Green-Lindsay}, which relies on (\ref{Fourier law})\
and two thermoelastic constitutive laws ($\varepsilon _{ij}=S_{ijkl}\sigma
_{kl}+A_{ij}\left( \Delta T+\tau _{1}\dot{T}\right) $ and $%
s=-T_{0}M_{ij}\varepsilon _{ij}+M_{ij}\left( \Delta T+\tau _{0}\dot{T}%
\right) $) where $s$ is the entropy density. The thermoelasticity field
equations of the first and third models are grasped in one system of two
coupled equations for the displacement-temperature pair:%
\begin{equation}
\begin{array}{c}
(C_{ijkl}u_{k},_{l}),_{j}+[M_{ij}(T+\tau _{1}\dot{T})],_{j}+b_{i}=\rho \ddot{%
u}_{i} \\ 
(k_{ij}T,_{j}),_{i}-c_{e}(\dot{T}+\tau _{0}\ddot{T})+\left( T_{0}M_{ij}\dot{u%
}_{i}\right) ,_{j}=-r.%
\end{array}%
\end{equation}%
The formulation of the G-L theory
implies $\tau _{1}>\tau _{0}\geq 0$ and, when both relaxation times are set
to zero, the C-C thermoelasticity is recovered.

$\allowbreak $In various situations (e.g., when the boundary conditions are
given in terms of stress tractions), it is advantageous to work with the
field equations expressed in terms of stresses. Then, an alternative
formulation of the L-S\ theory is obtained in terms of the stress-heat flux
pair $(\sigma _{ij},q_{i})$:

\begin{equation}
\begin{array}{c}
\left[ \rho ^{-1}\sigma _{(ik},_{k}\right] ,_{j)}+c_{\sigma }^{-1}\left(
A_{ij}\dot{q}_{k},_{k}-A_{ij}\dot{r}\right) +\left( \rho ^{-1}b_{(i}\right)
,_{j)} \\
=S_{ijkl}^{\prime }\ddot{\sigma}_{kl}, \\ 
\left[ c_{\sigma }^{-1}\left( q_{k},_{k}+r\right) \right] ,_{i}+T_{0}\left(
c_{\sigma }^{-1}A_{pq}\dot{\sigma}_{pq}\right) ,_{i}=-\lambda _{ij}{}\left( 
\dot{q}_{j}+\tau \ddot{q}_{j}\right) .%
\end{array}
\label{sigma-q pair}
\end{equation}%
Here $c_{\sigma }$ ($=c_{e}-T_{0}M_{ij}A_{ij}$) is the specific heat at zero
strain. Also, $\lambda _{ij}=k_{ij}^{-1}$ is the thermal resistivity tensor such that $%
\lambda _{ij}q_{i}q_{j}$ $>0$, and
\begin{equation}
S_{ijkl}^{^{\prime }}=S_{ijkl}-T_{0}c_{\sigma }^{-1}A_{ij}A_{kl},
\end{equation}%

The field equations of the G-L\ thermoelasticity in terms of the
stress-temperature pair $(\sigma _{ij},T)$ read \begingroup%
\renewcommand{\arraystretch}{1.5}%
\begin{equation}
\begin{array}{c}
\left[ \rho ^{-1}\sigma _{(ik},_{k}\right]
,_{j)}-A_{ij}t_{(0)}^{-1}[t_{1}c_{\sigma }^{-1}\left( k_{pq}\dot{T}%
,_{q}\right) ,_{p}-\left( \tau _{1}-\tau _{\left( 0\right) }\right) \ddot{T}]
\\ 
+\tilde{b}_{i},_{j}=\widetilde{S}_{ijkl}\ddot{\sigma}_{kl}, \\ 
c_{\sigma }^{-1}\left[ (k_{pq}T,_{q}),_{p}+r\right] -T_{0}c_{\sigma
}^{-1}A_{pq}\dot{\sigma}_{pq}=(\dot{T}+\tau _{(0)}\ddot{T}),%
\end{array}
\label{sigma-T pair}
\end{equation}%
\endgroup where we have $\widetilde{S}_{ijkl}=S_{ijkl}-\frac{\tau _{1}}{\tau _{(0)}}%
\frac{\theta _{0}}{c_{\sigma }}A_{ij}A_{kl}$, $\tilde{b}_{(ij)}=(\rho
^{-1}b_{(i}),_{j)}-\frac{\tau _{1}}{\tau _{(0)}}\frac{\dot{r}}{c_{\sigma }}%
A_{ij}$, and $\tau _{(0)}=\left( 1-\frac{c_{e}}{c_{\sigma }}\right) \tau _{1}+%
\frac{c_{e}}{c_{\sigma }}\tau _{2}$. The C-C theory is obtained from (\ref{sigma-q pair}) and (\ref{sigma-T pair}%
) by setting $\tau =0$ and $\tau _{1}=\tau _{0}=0$, respectively.

\bigskip

\section{From odd elasticity to odd compliance} As is well-known (e.g., 
\cite{Fruchart}), the major symmetry relation $C_{ijkl}=C_{klij}$ does not
hold in odd solids. To identify the elasticity ($C_{ijkl}$)\ and compliance (%
$S_{ijkl}$) tensors for an isotropic planar odd solid, we begin with%
\begin{equation}
\sigma _{kl}=K_{ijkl}u_{k},_{l},\text{ \ \ \ }i,j,k,l=1,2,
\label{odd-full-Hooke-law}
\end{equation}%
where $K_{ijkl}$\ is the tensor (with $\epsilon _{ij}$\ the Levi-Civita
symbol)%
\begin{equation}
\begin{array}{c}
K_{ijkl}=B\delta _{ij}\delta _{kl}-A\epsilon _{ij}\delta _{kl}+\mu \left(
\delta _{il}\delta _{jk}+\delta _{im}\delta _{jn}-\delta _{ij}\delta
_{kl}\right)  \\ 
+K^{0}\left( \epsilon _{ik}\delta _{jl}+\epsilon _{jl}\delta _{ik}\right) .%
\end{array}.  \label{K-tensor}
\end{equation}%
Hence, we find explicitly%
\begin{equation}
\begin{array}{c}
\sigma _{11}=\left( B+\mu \right)
u_{1},_{1}+K^{0}u_{1},_{2}+K^{0}u_{2},_{1}+\left( B-\mu \right) u_{2},_{2}
\\ 
\sigma _{12}=-\left( A+K^{0}\right) u_{1},_{1}+\mu u_{1},_{2}+\mu
u_{2},_{1}+\left( -A+K^{0}\right) u_{2},_{2} \\ 
\sigma _{21}=\left( A-K^{0}\right) u_{1},_{1}+\mu u_{1},_{2}+\mu
u_{2},_{1}+\left( -A+K^{0}\right) u_{2},_{2} \\ 
\sigma _{22}=\left( B-\mu \right)
u_{1},_{1}-K^{0}u_{1},_{2}-K^{0}u_{2},_{1}+\left( B+\mu \right) u_{2},_{2}%
\end{array}
\end{equation}%
from which we identify all $K_{ijkl}$'s.

Unfortunately $K_{ijkl}$\ is not invertible to a compliance form, so we cannot
write $u_{k},_{l}=K_{ijkl}^{-1}\sigma _{kl}$. In addition (\ref{odd-full-Hooke-law}%
)-(\ref{K-tensor}) imply\ $\sigma _{12}\neq \sigma _{21}$ which violates the
angular momentum balance under the assumption of no couple-stresses present,
suggesting that $A$ has to be removed to achieve invertibility. We then have $%
\ C_{ijkl}=K_{ijkl}$ (where $A=0$)  with the odd
elasticity property: $C_{1112}\neq C_{1211}$ and $C_{2212}\neq C_{1222}$.
Using $\sigma _{12}=\sigma _{21}$ we then write this Hooke's law in matrix
form\begingroup\renewcommand{\arraystretch}{1.9}%
\begin{equation}
\left( 
\begin{array}{c}
\sigma _{11} \\ 
\sigma _{22} \\ 
\sigma _{12}%
\end{array}%
\right) =\left[ 
\begin{array}{ccc}
B+\mu & B-\mu & K^{0} \\ 
B-\mu & B+\mu & -K^{0} \\ 
-K^{0} & K^{0} & \mu%
\end{array}%
\right] \left( 
\begin{array}{c}
e_{11} \\ 
e_{22} \\ 
2e_{12}%
\end{array}%
\right) ,  \label{odd-HL}
\end{equation}
\endgroup which specifies the elasticity matrix $\left[ \mathbf{C}\right] $
mapping the vector of elastic strains $\left( e_{11},e_{22},2e_{12}\right) $
into stresses, where elastic strain is defined as $e_{kl}=u_{(k},_{l)}$. The
compliance matrix $\left[ \mathbf{S}\right] =\left[ \mathbf{C}\right] ^{-1}$
is found$\allowbreak $ as\begingroup\renewcommand{\arraystretch}{1.9}%
\begin{equation}
\left[ \mathbf{S}\right] =\left[ 
\begin{array}{ccc}
\frac{\left( K^{0}\right) ^{2}+\mu ^{2}+B\mu }{4B\left( K^{0}\right)
^{2}+4B\mu ^{2}} & \frac{\left( K^{0}\right) ^{2}+\mu ^{2}-B\mu }{4B\left(
K^{0}\right) ^{2}+4B\mu ^{2}} & -\frac{K^{0}}{2\left( K^{0}\right) ^{2}+2\mu
^{2}} \\ 
\frac{\left( K^{0}\right) ^{2}+\mu ^{2}-B\mu }{4BK^{2}+4B\mu ^{2}} & \frac{%
\left( K^{0}\right) ^{2}+\mu ^{2}+B\mu }{4BK^{2}+4B\mu ^{2}} & \frac{K^{0}}{%
2\left( K^{0}\right) ^{2}+2\mu ^{2}} \\ 
\frac{K^{0}}{2\left( K^{0}\right) ^{2}+2\mu ^{2}} & -\frac{K^{0}}{2\left(
K^{0}\right) ^{2}+2\mu ^{2}} & \frac{\mu }{\left( K^{0}\right) ^{2}+\mu ^{2}}%
\end{array}%
\right] \allowbreak \allowbreak \allowbreak .  \label{S-matrix}
\end{equation}%
\begingroup\renewcommand{\arraystretch}{1.9}\endgroup Here we identify the
planar bulk compliance $B^{-1}$ and the shear compliance $\allowbreak
S_{1212}=\mu /\left( \left( K^{0}\right) ^{2}+\mu ^{2}\right) $.

In analogy to the 3d case (\cite{Mura}, \cite{Jasiuk-Boccara}), to deal with
internal stresses and strains in odd thermoelasticity, we consider the total
strain $\varepsilon _{ij}$ in the body to be the sum of the elastic strain $%
e_{ij}$ given above and the eigenstrain $\varepsilon _{ij}^{\ast }$ 
\begin{equation}
\begin{array}{c}
\varepsilon _{ij}=e_{ij}+\varepsilon _{ij}^{\ast }\text{ \ where} \\ 
e_{ij}=S_{ijkl}\sigma _{kl}\text{ \ and \ }\varepsilon _{ij}^{\ast
}=A_{ij}\Delta T\text{\ ,\ \ }i,j=1,2.%
\end{array}
\end{equation}%
The eigenstrain $\varepsilon _{ij}^{\ast }$ contributes to the linear
momentum balance as a body force, and to the boundary conditions as a
surface force. In general, eigenstrains and eigenstresses can also be due to
swelling, plastic or transformation, loss/gain of mass, or
changes to the molecular structure of the phases.

\section{Stress field invariance under a shift in compliances}

In the realm of materials science and mechanical engineering, the study of invariant properties of stress in plane elasticity and equivalence classes of composites plays a pivotal role. Plane elasticity simplifies stress and deformation to a two-dimensional perspective, essential for analyzing thin plates or surfaces under load. The invariants of stress, quantities unaltered by coordinate transformations, are crucial for comprehending material behavior under varying loads, aiding in predicting material failure and other mechanical properties. Concurrently, composites, engineered from multiple materials with distinct properties, are classified into equivalence classes based on shared characteristics such as mechanical or thermal behavior. This classification is vital for material selection in engineering, enabling the choice of composites that meet specific requirements such as strength, weight, or heat resistance. Together, these concepts are fundamental in optimizing material properties for varied applications, ensuring safety and efficiency in high-stress environments or when employing advanced materials. Understanding stress invariance in the context of active odd solids can provide insights into their mechanical properties and stability, guiding their integration into engineering and material science applications where their unique characteristics might be advantageous.

We now take the odd elastic body to occupy a simply-connected domain $%
\mathcal{B}$ in the plane, with a boundary $\partial B$ characterized by the
unit outer normal vector $n_{i}$. The body is assumed to be in static
equilibrium ($\sigma _{ij},_{j}=0$) while subjected to traction boundary
conditions on its entire boundary ($\sigma _{ij}n_{j}=t_{i}^{\left( n\right)
}$, $\forall x_{i}\in \partial B$), and to satisfy the global equilibrium 
\begin{equation}
\int_{\partial B}t_{i}^{\left( n\right) }dS=0\text{ \ \ \ \ }\int_{\partial
B}\epsilon _{ijk}x_{j}t_{k}^{\left( n\right) }dS=0.
\label{global equilibrium}
\end{equation}%
If the body domain is multiply-connected, the tractions are
self-equilibrated (with overall zero force and zero moment) on each internal
boundary. We take the compliances and eigenstrains to be, in general,
inhomogeneous in $\mathcal{B}$ and assume them twice-differentiable in $%
\mathcal{B}$.

The stress invariance problem \cite{CLM} is described by asking the
following question: "\textit{Given a statically equilibrated solid with a
stress field }$\mathbf{\sigma }=\left( \sigma _{11},\sigma _{22},\sigma
_{12}\right) $\textit{\ under prescribed traction boundary conditions, can
the compliance tensor }$S_{ijkl}$\textit{\ be changed to a new }$\widehat{S}%
_{ijkl}$\textit{\ in such a way that the stress field remains unchanged}?"
Now, the strain compatibility condition $\varepsilon _{11,22}+\varepsilon
_{22,11}=2\varepsilon _{12,12}$ becomes%
\begin{equation}
\begin{array}{c}
\nabla ^{2}\left[ \left( B^{-1}+S_{1212}\right) \left( \sigma _{11}+\sigma
_{22}\right) \right]  \\ 
-2\left( S_{1212},_{11}\sigma _{11}+2S_{1212},_{12}\sigma
_{12}+S_{1212},_{22}\sigma _{22}\right)  \\ 
=4[\left( S_{1211}\sigma _{11}\right) ,_{12}+\left( S_{1222}\sigma
_{22}\right) ,_{12}-\left( S_{1211}\sigma _{12}\right) ,_{22} \\ 
-\left( S_{1212}\sigma _{12}\right) ,_{11}]+8\varepsilon _{12,12}^{\ast
}-4\varepsilon _{11,22}^{\ast }-4\varepsilon _{22,11}^{\ast },%
\end{array}
\label{2d-compatibility}
\end{equation}%

Inspecting (\ref{2d-compatibility}) we see that, for the new stress field to
remain $\sigma _{ij}$, these relations must hold%
\begin{equation}
\begin{array}{c}
\widehat{B}^{-1}+\widehat{S}_{1212}=m\left( B^{-1}+S_{1212}\right) \text{ \
\ \ }\widehat{S}_{1212,11}=mS_{1212},_{11} \\ 
\widehat{S}_{1212},_{12}=mS_{1212},_{12}\text{\ \ \ \ }\widehat{S}%
_{1212},_{22}=mS_{1212},_{12},%
\end{array}
\end{equation}
where $m$ is an arbitrary scalar. This implies%
\begin{equation}
\widehat{B}^{-1}=mB^{-1}+a+bx_{1}+cx_{2},\text{ \ }\widehat{S}%
_{1212}=mS_{1212}-a-bx_{1}-cx_{2}.
\end{equation}%
The constants $m$, $a$, $b$, and $c$ are subject to restrictions dictating
that the new compliances be non-negative.

In other words, although $S_{ijkl}\neq S_{klij}$, the answer to the above
question is affirmative for a so-called \textit{shift} of $S_{ijkl}$ to $%
\widehat{S}_{ijkl}$\ according to \ 
\begin{equation}
\widehat{S}_{ijkl}=S_{ijkl}+S_{ijkl}^{I},  \label{CLM-shift}
\end{equation}%
where%
\begin{equation}
S_{ijkl}^{I}\left( \Lambda ,-\Lambda \right) =\frac{1}{2\Lambda }\delta
_{ij}\delta _{kl}-\frac{1}{4\Lambda }\left( \delta _{ik}\delta _{jl}+\delta
_{il}\delta _{jk}\right) .  \label{CLM-shift tensor}
\end{equation}%
is the \textit{shift tensor} and $\Lambda ^{-1}$\ is linear in $x_{1}$ and $%
x_{2}$, only subject to a condition that the new stiffness tensor $\widehat{S%
}_{ijkl}$ is positive-definite everywhere in $\mathcal{B}$.

Writing this in terms of matrices, the shift (\ref{CLM-shift}) with (\ref%
{CLM-shift tensor})\ is expressed as a change of the compliance matrix $%
\left[ \mathbf{S}\right] $ to a new $\left[ \widehat{\mathbf{S}}\right] $\
according to%
\begin{equation}
\left[ \widehat{\mathbf{S}}\right] =\left[ \mathbf{S}\right] +\left[ \mathbf{%
S}^{I}\right] ,\text{ \ }\left[ \mathbf{S}^{I}\right] =\frac{1}{2\Lambda }%
\begin{bmatrix}
0 & 1 & 0  \\
1 & 0 & 0 \\
0 & 0 & -2%
\end{bmatrix}%
\label{CLM-shift matrix}
\end{equation}%
Thus, a very wide range of odd thermoelastic materials with different values
of material properties will have the same new stress field $\widehat{\mathbf{%
\sigma }}$ as the original $\mathbf{\sigma }$.

\section{Hydrodynamic modes}
As an instructive example of collective phenomena in thermoelastic models, we study collective modes in the following model of thermoelasticity
\begin{equation}\label{eq:theromelc}
\begin{array}{c}
(C_{ijkl}u_{k},_{l}),_{j}+\left( M_{ij}T\right) ,_{j}=\rho \ddot{u}_{i}
\\ 
\left( k_{ij}T,_{j}\right) ,_{i}
+T_{0}\left[ M_{ij}\dot{u}_{i}\right] ,_{j}=c_{e}\dot{T},%
\end{array}
\end{equation}%
\bigskip
where $k_{ij}=k_1\delta_{ij}+k_2 \epsilon_{ij}$ and $m_{ij}=m_1\delta_{ij}+m_2 \epsilon_{ij}$. 

The above system of equation can be put in the hydrodynamic form by introducing an auxiliary velocity field $v=\dot u$ that plays the role of a Josephson equation in a system with a spontaneously broken translation symmetry. It is convenient to pass to the Fourier space, where the excitations are proportional to plane waves $e^{i(k\cdot x-\omega t)}$. Our aim is to compute the dispersion relations $\omega=\omega(k)$ (see e.g. \cite{Martin}). 


We get a fifth-order equation that does not have a closed-form solution. However, for our purposes, it is enough to construct a perturbative solution in the powers of $k$ as $k\rightarrow0$. We expect two pairs of sound modes plus a diffusive mode due to the temperature profile. Therefore our perturbative ansatz for the solution reads 
$\left(\omega^2 + 
   i \omega k^2 \Gamma_1(\omega, k)  - 
   v_1^2(\omega, k) k^2 \right)\times 
   \left(\omega^2 + 
   i \omega k^2 \Gamma_2(\omega, k) - 
   v_2^2(\omega, k) k^2\right) 
   \left(-i \omega + 
   k^2 D(\omega, k)\right).$
In the lowest order of expansion $\Gamma_1(\omega, k) =\Gamma_1 $, $\Gamma_2(\omega, k) =\Gamma_2$, $D(\omega, k) =D$. These coefficients correspond to the sound attenuation and the diffusion coefficients respectively. $v_1^2(\omega, k)$ and $v_2^2(\omega, k) $ correspond to non-dissipative sound velocities squared. The solution reads
\begin{widetext}
\begin{subequations}
\begin{equation}
\begin{split}
v_1^2&= \frac{1}{\rho^2}[ -(B c_e + (m_1^2 - m_2^2 )T_0 + 2 c_e \mu) \rho \\
+&\sqrt{(8 m_1 m_2 T_0 K^0 + 4 m_1^2 T_0 \mu - 
   4 m_2^2 T_0 (B + \mu) + (B c_e + (m_1^2 - m_2^2) T_0 + 
     2 c_e \mu)^2 + 4 c_e ((K^0)^2 + \mu (B + \mu))) \rho^2} ].
\end{split}
\end{equation}
\begin{equation}
\begin{split}
v_2^2&= \frac{1}{\rho^2}[ -(B c_e + (m_1^2 - m_2^2 )T_0 + 2 c_e \mu) \rho \\
-&\sqrt{(8 m_1 m_2 T_0 K^{0} + 4 m_1^2 T_0 \mu - 
   4 m_2^2 T_0 (B + \mu) + (B c_e + (m_1^2 - m_2^2) T_0 + 
     2 c_e \mu)^2 + 4 c_e ((K^0)^2 + \mu (B + \mu))) \rho^2} ].
\end{split}
\end{equation}
\end{subequations}
\end{widetext}
 Similarly, we can determine the sound attenuation coefficients
 \begin{subequations}
\begin{equation}
\Gamma_1= -2\frac{2 k_1 (4 (K^0)^2 + (2 \mu - v_1^2 \rho) (2 B + 2 \mu - 
      v_1^2 \rho))}{v_1^2  (v_1^2  - v_2^2 ) \rho^2},
\end{equation}
\begin{equation}
\Gamma_2= -2\frac{2 k_1 (4 (K^0)^2 + (2 \mu - v_2^2 \rho) (2 B + 2 \mu - 
      v_2^2 \rho))}{v_2^2  (v_2^2  - v_1^2 ) \rho^2},
\end{equation}
\end{subequations}
and the diffusion coefficient
\begin{equation}
D=\frac{2 k_1 (4 (K^0)^2 + (2 \mu - v_1^2 \rho) (2 B + 2 \mu - v_1^2 \rho))}{v_1^2 (v_1^2 - v_2^2) \rho^2}.
\end{equation}
We note that generically in models of thermoelasticity without odd coefficients only dilatational waves are affected by thermal effects (see e.g. \cite{li_dispersion_2021}). Here odd coefficients mix transverse and longitudinal waves, which means that shear waves also feel temperature profiles. Given the current experimental effort to understand odd solids (see e.g. \cite{tan_odd_2022}), our study suggests that temperature measurements can shed light on odd transport coefficients. In addition, odd coefficients also affect the dissipative responses as they modify both sound attenuation and diffusion.

\section{Discussion}

In this work we have developed the framework of odd thermoelasticty and investigated several properties of odd elastic solids. The role of thermal effects in odd elasticty has been limited to the Fourier heat conduction. In this paper we show, using the framework of extended irreversible thermodynamics, that in active solids it has to be modified to the odd generalization of the Maxwell-Cattaneo heat conduction. Building upon this observation we propose odd theories of thermoelasticty that should be applicable to odd solids under various conditions.

Motivated by the potential applications of active solids, in the subsequent part of the manuscript we investigate the stress formulation of odd elasticity. We derive the odd compliance matrix and show that the stress is invariant under specific shift of compliance matrix coefficients.

Finally we investigate collective modes in a model that exhibits odd elasticity and violates Onsager reciprocity relations. The main finding of this analysis is that odd transport coefficients manifest themselves in temperature profiles. Our analysis paves the way for a systematic study of thermal effects in odd solids that can also play an important role in micropolar or viscoelastic solids.

\section*{Acknowledgments}
 M.O.-S. thanks the Isaac Newton Institute for Mathematical Sciences, University of Cambridge, for support and hospitality as the Rothschild Distinguished Visiting Fellow during the program "Uncertainty quantification and stochastic modelling of materials" where work on this paper was partially completed. P.S. acknowledges support from the Polish National Science Centre (NCN) Sonata Bis grant 2019/34/E/ST3/00405. Part of this work was performed at the Aspen Center for Physics, which is supported by the National Science Foundation grant PHY-2210452.

\bibliography{Bibliography.bib}

\end{document}